\documentclass[preprintnumbers]{revtex4}
\usepackage{graphicx}
\usepackage{color}
\usepackage{eurosym}
\usepackage{amssymb}
\usepackage{amsmath}
\usepackage{amsfonts}

\input{tcilatex}
\begin{document}

\title{Lightest Kaonic Nuclear Clusters}
\author{Roman Ya. Kezerashvili$^{1,2}$\thanks{%
Talk presented at the Twelfth Conference on the Intersections of Particle
and Nuclear Physics, CIPANP 2015.}, Shalva M. Tsiklauri$^{3}$, Nurgali Zh.
Takibayev$^{4}$}
\affiliation{\mbox{$^{1}$Physics Department, New
York City College of Technology, The
City University of New York,} \\
Brooklyn, NY 11201, USA \\
\mbox{$^{2}$The Graduate School and University Center, The
City University of New York,} \\
New York, NY 10016, USA \\
\mbox{$^{3}$Borough of Manhattan Community College, The City University of
New York,}\\
New York, NY 10007, USA \\
\mbox{$^{4}$Al-Farabi Kazakh National University, 480078, Almaty, Kazakhstan
}}

\begin{abstract}
We present our study of kaonic three-body $\overset{\_}{K}NN$, $\overset{\_}{%
K}\overset{\_}{K}N$ and $KK\overset{\_}{K}$ and four-body $\overset{\_}{K}%
NNN $, $\ $and $\overset{\_}{K}\overset{\_}{K}NN$ clusters within the
framework of a potential model using the method of hyperspherical functions
in momentum representation. To perform a numerical calculations for the
bound state energy of the light kaonic system, we use a set of different
potentials for the nucleon-nucleon and $\overset{\_}{K}N$ interactions, as
well as for the kaon-kaon interaction. The calculations show that a
quasibound state energy is not sensitive to the $NN$ interaction, and it
shows very strong dependence on the $\overset{\_}{K}N$ potential. We also
compare our results with those obtained using different theoretical
approaches. The theoretical discrepancies in the binding energy and width
for the lightest kaonic system related to the different $NN$ and $\overset{\_%
}{K}N$ interactions are addressed.
\end{abstract}

\maketitle


\section{Introduction}

\bigskip Nowadays, the study of exotic nuclear systems involving a $\overset{%
\_}{K}$ is an important topic in hadron physics, because the existence of
kaonic nuclear states is related to kaon condensation and to physics of the
core of neutron stars that by today's understanding are built up from exotic
matter: pion and kaon condensates and quark matter \cite{Brown1, Lee1996}.
Kaonic nuclei carry important information concerning the $\overset{\_}{K}-$%
nucleon interaction in the nuclear medium. This information is very
important in understanding kaon properties at finite density and in
determining of the constraints on kaon condensation in high-density matter.
The latter will allow one to adjust the methods developed in condensed
matter physics for exciton and excitonic polariton condensates (see, for
example, \cite{BermanKezPRB2012Exciton, BermanKezPRB2012Polariton}) to study
the kaon condensation. The best way to understand the many body kaonic
nuclear system is to study the simplest two-, three- and four-body clusters: 
$\overset{\_}{K}N$ , $\overset{\_}{K}NN,$ and $\overset{\_}{K}NNN$, as well
as double kaonic clusters, when one nucleon in the three- or four-body
kaonic cluster is replaced by the $K-$meson. The light kaonic clusters $%
\overset{\_}{K}NN$ and $\overset{\_}{K}NNN$ represent three- and four-body
systems and theoretically can be treated within the framework of a few-body
physics approaches. In the recent past much efforts have been focused on the
calculations of quasibound state energies and widths for three- and
four-body kaonic clusters. A variety of methods have been used in
configuration and momentum spaces, to obtain eigenvalues for energy and
width of quasibound states using diverse sets of $\overset{\_}{K}N$ and $NN$
interactions. These include but are not limited by variational method
approaches \cite{AYPRC2002, AYPL2002, AYPL2004, ADYPL2005, DotePRC2004,
AYKN, DoteProg2002, DW2007, DHWnp2008, DHW, WycechGreen}, the method of
Faddeev equations in momentum and configuration spaces \cite{Nina1, Nina2,
IkedaSato1, IkedaSato2, Oset, IkedaKamanoSato3, FaddFixCent1, FaddFixCent2,
FaddFixCent3, FaddFixCent4, FadYa4body, Nina2014, Kezerashvili2015},
Faddeev-Yakubovsky equations \cite{FadYa4body} and the method of
hyperspherical harmonics in configuration and momentum spaces \cite{BarneaK,
RKSh.Ts2014, Kezerashvili2015}. However, the predicted values for the
binding energy and the width are in considerable disagreement. For example,
for the $K^{-}pp$ cluster the predicted values for the binding energy and
the width are 9--95 MeV and 20--110 MeV, respectively.

On the experimental side, several experiments have been performed to search
for the kaonic clusters using various nuclear reactions starting from the
first measurement reported by the FINUDA collaboration for the $K^{-}pp$
cluster \cite{AgnelloKpp} and including the most recent report of J-PARC E15
Collaboration \ \cite{JPARCE15N}. The situation is still controversial and
the existence, for example, of the $K^{-}pp$ quasibound state has not been
established yet. Thus, the theoretical and experimental study of composite
systems of $K-$mesons and nucleons is still a challenging issue in nuclear
physics.

Below we present a study of the lightest kaonic nuclear clusters using the
method of hyperspherical functions. We focus on three- and four-body
nonrelativistic calculations within the framework of a potential model for
the three- and four-body kaonic clusters using the method of hyperspherical
harmonics (HH) in momentum representation. Calculations for a binding energy
and width of the kaonic three- and four-body system are performed using
different $NN$ potentials and kaon-nucleon interactions, as well as
kaon-kaon interactions. Such approach allows one to understand the key role
of the kaon-nucleon interaction and the importance of nucleon-nucleon
interaction in the formation of quasibound states of the kaonic three- and
four-body systems.

\section{Theoretical formalism}

The Hamiltonians of the three and four nonrelativistic particles for the $%
\overset{\_}{K}NN$ and $\overset{\_}{K}NNN$ systems, respectively, read as%
\begin{equation}
H_{3}=\widehat{T}_{3}+V_{N_{1}N_{2}}+V_{\overset{\_}{K}N_{1}}+V_{\overset{\_}%
{K}N_{2}},  \label{Hamilton}
\end{equation}

\begin{equation}
H_{4}=\widehat{T}_{4}+\sum\limits_{1\leq i<j\leq
3}V_{N_{i}N_{j}}+\sum\limits_{i=1}^{3}V_{\overset{\_}{K}N_{i}},
\label{Hamilton4}
\end{equation}%
where $\widehat{T}_{3}$ \ and $\widehat{T}_{4}$ \ are the operators of the
kinetic energy for three- and four-particle system, respectively, $%
V_{N_{i}N_{j}}$ is the nucleon-nucleon potential and $V_{\overset{\_}{K}%
N_{i}}$ is a pairwise effective antikaon interaction with the nucleon. The
effective interactions of the $\overset{\_}{K}N,$ $KN,$ $\overset{\_}{K}%
\overset{\_}{K}$ and $K\overset{\_}{K}$ two-body subsystems are discussed in
detail in Refs. \cite{AYPRC2002, AYPL2002, AYKN, HW, DHW, Weise, JidoKanada,
KanadaJido}. Below, we use two effective $\overset{\_}{K}N$ interactions
that were derived in different ways. The effective $\overset{\_}{K}N$
interactions can be derived phenomenologically or constructed using the
chiral SU(3) effective field theory, which identifies the Tomozawa-Weinberg
terms as the main contribution to the low-energy $\overset{\_}{K}N$
interaction \cite{Weise}. The single-channel potential for the description
of the $\overset{\_}{K}N$ interaction was derived in Refs. \cite{AYPRC2002,
AYKN} phenomenologically, using $\overset{\_}{K}N$ scattering and kaonic
hydrogen data and reproducing the $\Lambda (1405)$ resonance as a $K^{-}p$
bound state at 1405 MeV, and the decay width of Lambda(1405) is also taken
into account in this potential.We refer to this as the Akaishi-Yamazaki (AY)
potential. The AY potential is energy independent. The other $\overset{\_}{K}%
N$ interaction given in Ref. \cite{HW} was derived based on the chiral
unitary approach for the $s-$wave scattering amplitude with strangeness $%
S=-1 $, and reproduces the total cross sections for the elastic and
inelastic $K^{-}p$ scattering, threshold branching ratios, and the $\pi
\Sigma $ mass spectrum associated with the $\Lambda $(1405). Hereafter we
refer to this energy-dependent potential for the parametrization \cite%
{hwHWJH} as the HW potential. Both potentials are constructed in the
coordinate space, are local, and can be written in the one-range Gaussian
form as

\begin{equation}
V_{\overset{\_}{K}N}(r)=\sum\limits_{I=0,1}U^{I}\exp \left[ -\left(
r/b\right) ^{2}\right] P_{\overset{\_}{K}N}^{I},  \label{KNpotential}
\end{equation}%
where $r$ is the distance between the kaon and the nucleon, $b$ is the range
parameter and $P_{\overset{\_}{K}N}^{I}$ is the isospin projection operator$%
. $ The values of the potential depth $U^{I=0}$ and $U^{I=1}$ for each \
interaction are given in Refs. \cite{AYPRC2002, AYKN} and \cite{hwHWJH} and
the range parameter is chosen to be $b=0.66$ fm for the AY potential and $%
b=0.47$ fm for the HW potential.

To describe the $V_{N_{i}N_{j}}$ nucleon-nucleon interaction, we use several
different $NN$ potentials: the realistic Argonne V14 (AV14) and V18 (AV18) 
\cite{ArgonneV14, ArgonneV18}, the semi-realistic Malfliet and Tjon MT-I-III
(MT) \cite{MT} potential, the Tamagaki G3RS potential \cite{Tpotential},
which we hereafter refer to as the T potential, and potential \cite{Minesota}%
, which we refer to as the M potential. Therefore, the use of different $NN$
potentials and $\overset{\_}{K}N$ interactions allows one to perform a
validity test for the lightest kaonic clusters against various $NN$ and $%
\overset{\_}{K}N$ interactions.

The binding energies and the wave functions of the three and four
nonrelativistic particle can be obtained by solving the Schr\"{o}dinger
equation with the Hamiltonians (\ref{Hamilton}) and (\ref{Hamilton4}),
respectively. In our approach we use the hyperspherical harmonics method
that represents a technique of solution of the Schr\"{o}dinger equation to
find the bound and scattering states for a few body system. The main idea of
this method is the expansion of the wave function of the corresponding
nuclear system in terms of \ hyperspherical harmonics that are the
eigenfunctions of the angular part of the Laplace operator in the
six-dimensional space (three-body problem) or in the nine-dimensional space
(four-body problem).\ The details of this method can be found in the
monographs \cite{DzhbutiKrup, Avery, JibutiSh}. In our calculations we use
the HH method in momentum representation \cite{JibutiNuclPhys, JibutiSh}.
One starts from the Schr\"{o}dinger equation for the three or four particles
with the Hamiltonians (\ref{Hamilton}) and (\ref{Hamilton4}), respectively,
and rewrites this equation in the integral form in the momentum
representation using the set of the Jacobi momenta $\mathbf{q}_{i}$ in $%
3(N-1)$-dimensional momentum space. These momenta are the trees of Jacobi
coordinates\ for three- or four-particle system $\mathbf{q}_{i}=\sqrt{\frac{%
m_{12...i}m_{i+1}}{m_{12...i+1}}}\left( \frac{1}{m_{12...i}}%
\sum\limits_{j=1}^{i}m_{j}\mathbf{q}_{j}-\mathbf{q}_{i+1}\right) ,$ \ $%
i=1,2,...,N-1,$ where $m_{j}$ and $\mathbf{q}_{j}$ are the particles masses
and momentum vectors conjugated to the position vectors $\mathbf{r}_{j}$
respectively, $m_{12...i}=\sum\limits_{j=1}^{i}m_{j}$ and $N$ is the number
of particles. After that, one introduces the set of the hyperspherical
coordinates in the momentum space given by the hyperradius $\varkappa ^{2}$= 
$\sum\limits_{i=1}^{N-1}q_{i}^{2}$ and the set of angles $\Omega _{\varkappa
}$, which define the direction of the vector $\mathbf{\varkappa } $ in $%
3(N-1)$-dimensional momentum space, as well as the symmetrized
hyperspherical harmonics in momentum representation $\Phi _{\mu }^{\lambda
}(\Omega _{\varkappa },\mathbf{\sigma },\mathbf{\tau })$ that are written as
a sum of products of spin and isospin functions and hyperspherical harmonics 
\cite{JRK}. Above, for the sake of simplicity, we denoted by $\lambda $ the
totality of quantum numbers on which the $N-$body hyperspherical harmonics
depend and the integer $\mu $ is the global momentum in the $3(N-1)$%
-dimensional configuration space, which is the analog of angular momentum in
case of $N=2$. The HH are the eigenfunctions of the angular part of the $%
3(N-1)$-dimensional Laplace operator in configuration space with eigenvalue $%
L_{N}(L_{N}+1)$, where $L_{N}$ $=\mu +3(N-2)/2$ and they are expressible in
terms of spherical harmonics and Jacobi polynomials \cite{DzhbutiKrup,
Avery, JibutiSh}. By expanding the wave function of $N$ bound particles in
terms of the symmetrized hyperspherical harmonics in momentum space

\begin{equation}
\Psi (\varkappa ,\Omega _{\varkappa })=\varkappa ^{-\frac{3N-4}{2}%
}\sum\limits_{\mu ,\lambda }u_{\mu }^{\lambda }(\varkappa )\Phi _{\mu
}^{\lambda }(\Omega _{\varkappa },\overrightarrow{\sigma },\overrightarrow{%
\tau }),  \label{ExpanGeneral}
\end{equation}%
and substituting Eq. (\ref{ExpanGeneral}) into the corresponding integral
Schr\"{o}dinger equation in the momentum representation, one obtains a
system of coupled integral equations for the hyperradial functions\ $u_{\mu
}^{\lambda }(\varkappa )$ for the system of three or four particles.\ The
detailed description of the formalism for the $K^{-}pp$ cluster can be found
in Ref. \cite{Kezerashvili2015}. Here we expand the wave function of three
bound particles in terms of the symmetrized hyperspherical harmonics $\Phi
_{\mu }^{l_{p}l_{q}L}(\Omega _{\varkappa },\mathbf{\sigma },\mathbf{\tau })$
in momentum representation: $\Psi (\mathbf{p},\mathbf{q})=\sum\limits_{\mu
l_{p}l_{q}}u_{\mu }^{l_{p}l_{q}L}(\varkappa )\Phi _{\mu
}^{l_{p}l_{q}L}(\Omega _{\varkappa },\mathbf{\sigma },\mathbf{\tau }),$
where $\mu $ is the grand angular momentum, $L$ is the total orbital
momentum, $l_{p}$ and $l_{q}$ are the angular momenta corresponding to the
Jacobi momenta $p$ and $q$ that are conjugated to the standard Jacobi
coordinates for three particles, $\varkappa $ is the hyperradius in the six
dimensional momentum space$,$ and $\Omega _{\varkappa }$ is the set of five
\ angles which define the direction of the vector $\mathbf{\varkappa }.$ The
functions $\Phi _{\mu }^{l_{p}l_{q}L}(\Omega _{\varkappa },\mathbf{\sigma },%
\mathbf{\tau })$ are written as a sum of products of spin and isospin
functions and HH, using the Raynal-R\'{e}vai coefficients \cite{RaynalRevai}%
. For the system $K^{-}pp$ the wave function is antisymmetrized with respect
to two protons, while for the $K^{-}K^{-}p$ system it is symmetrized with
respect to two antikaons. For the hyperradial functions $u_{_{\mu
}}^{l_{q}l_{p}L}(\varkappa )$ we obtain the coupled integral equations. By
solving the coupled integral equations one can find the hyperradial
functions $u_{_{\mu }}^{l_{q}l_{p}L}(\varkappa )$ for a given $L$ and the
binding energies for the $K^{-}pp$ and $K^{-}K^{-}p$ systems. For the system 
$\overset{\_}{K}NNN$ the wave function is antisymmetrized with respect to
three nucleons, while for the $K^{-}K^{-}pp$ system it is symmetrized with
respect to two antikaons and antisymmetrized with respect to two protons.

\section{Results of numerical calculations and discussion}

\subsection{$K^{-}pp$ cluster}

\bigskip\ Let's start with the results of our calculations of the $K^{-}pp$
cluster recently reported in Ref. \cite{Kezerashvili2015}. Results of these
calculations for the $K^{-}pp$ cluster are presented in Table~\ref{tab1N}.
For the calculations of the binding energy and the width with the method of
HH we use as input MT, T, and AV14 potentials for the $NN$ interaction,
while for the $\overset{\_}{K}N$ interaction we use the energy-dependent
effective HW and the phenomenological AY potentials. Such an approach
allowed us to examine how the $K^{-}pp$ cluster's structure depends on
different choices of the $\overset{\_}{K}N$ interactions for the same $NN$
potential, as well as to investigate its dependence on different choices of
the $NN$ interaction for the same $\overset{\_}{K}N$ interaction, and to
understand the sensitivity of the system to the input interactions.\emph{\ }%
The analysis of the calculations presented in Table \ref{tab1N} show that
the AY potential as the $\overset{\_}{K}N$ interaction input falls into the $%
46-47$ MeV range for\emph{\ }the binding energy\emph{\ }of the $K^{-}pp$
cluster, while the chiral HW $\overset{\_}{K}N$ potential gives about $%
17-21.6$ MeV for the binding energy. Thus, the values for the binding energy
for the $K^{-}pp$ system obtained for the different $NN$ potentials are in
reasonable agreement, and the ground state energy is not very sensitive to
the $NN$ interaction. However, there is a very strong dependence on the
antikaon-nucleon interaction. When we employ the effective energy-dependent
chiral-theory based HW potential for the $\overset{\_}{K}N$ interaction and
different $NN$ interactions, as inputs, we predict a weakly bound $K^{-}pp$
cluster. This is similar to Ref. \cite{DHW}, where the authors employed
several versions of energy-dependent effective $\overset{\_}{K}N$
interactions derived from chiral SU(3) dynamics together with the realistic
AV18 $NN$ potential. Our calculations also confirm results reported in
earlier studies employing the same type of $\overset{\_}{K}N$ interaction 
\cite{DHWnp2008, IkedaKamanoSato3, BarneaK}. The energy of the bound state,
as well as the width calculated for the AY potential \cite{AYPRC2002} are
more than twice as big as those obtained for the energy-dependent chiral $%
\overset{\_}{K}N$ HW potential. Therefore, the highest binding energies are
obtained for the phenomenological AY potential. Let's compare our results
with those obtained with different variational approaches. Our results are
in good agreement with those of Ref. \cite{AYKN}, where the binding energy
and the width for the $K^{-}pp$ cluster were calculated by employing the AY
potential as the $\overset{\_}{K}N$ interaction and T potential as the $NN$
interaction. Recently it was reported that "resonance and coupled-channel
problem are key ingredients in the theoretical study of the $K^{-}pp$" \cite%
{FeshbaxRes2014}$.$ Those authors employ a coupled-channel Complex Scaling
Method combined with the Feshbach method since this approach can
simultaneously treat these two ingredients. Interestingly enough, their
calculations \cite{FeshbaxRes2014, FeshbaxRes2015} for the binding energy
and width are consistent with our results obtained within the single channel
potential model. The comparison of our calculations with results obtained
using the HH method in configuration space \cite{BarneaK} and differential
Faddeev equations \cite{Kezerashvili2015} also are in reasonable agreement.
This is a good indication that the binding energy does not depend
significantly on the method of calculation.

\begin{table}[tbp]
\caption{The binding energy $B$ and width $\Gamma $ for the $K^{-}pp$ system
calculated in the framework of the method of HH in the momentum
representation for different interactions. $NN$ potentials: AV14 
\protect\cite{ArgonneV14}, MT \protect\cite{MT} and T \protect\cite%
{Tpotential}. $\protect\overset{\_}{K}N$ interactions: AY \protect\cite%
{AYPRC2002} and HW \protect\cite{HW}. $E_{K^{-}p}$ is two-body energy of the 
$K^{-}pp$ cluster.}
\label{tab1N}\centering
\begin{tabular}{c|ccc|c|c|c}
\hline\hline
& AV14+AY & MT+AY & T+AY & AV14+HW & MT+HW & T+HW \\ \hline
$B$, MeV & 46.2 & 46.5 & 46.3 & 17.2 & 20.5 & 20.6 \\ 
$\Gamma$, MeV & 66.7 & 84.3 & 74.5 & 44.3 & 48.1 & 49.5 \\ \hline
$E_{K^{-}p},$ MeV & \multicolumn{3}{|c|}{29.9} & \multicolumn{3}{|c}{10.9}
\\ \hline\hline
\end{tabular}%
\end{table}

\subsection{$K^{-}K^{-}p$ \textit{cluster}}

Three-body problem with two mesons and one baryon have received considerable
attention in the recent literature \cite{JidoKanada, KanadaJido, OsetPiPiN,
OsetThreeBodyhadron}. The baryonic systems $\overset{\_}{K}\overset{\_}{K}N$
and $\overset{\_}{K}KN$ with two kaons were investigated in Refs. \cite%
{JidoKanada, KanadaJido, Judo2010}. We study a possible bound state of the $%
K^{-}K^{-}p$ cluster with $S=-2$, $I=1/2,$ $J^{+}=1/2^{+}$ using the
effective $s-$wave AY and HW potentials assuming that this state is formed
due to the strong $K^{-}p$ attraction. The strength of the $s-$wave $\overset%
{\_}{K}\overset{\_}{K}$ interaction for the isospin $I=0$ is zero due to
Bose statistics, and we consider a weak repulsion for the isospin $I=1$ that
reproduces the scattering lengths $a_{K^{+}K^{+}}=-0.14$~fm for the range
parameter $b=0.66$~fm (AY potential ) and $b=0.47$~fm (HW potential). The
results of calculations for the binding energies for the $K^{-}p$ and $%
K^{-}K^{-}p$, the bound $K^{-}K^{-}p$ state without $K^{-}K^{-}$
interaction, and the root-mean-square radius of the $\overset{\_}{K}$
distribution are presented in Table~\ref{tab2}. For the AY potential, the $%
K^{-}K^{-}p$ system is still bound even with a much stronger $K^{-}K^{-}$
repulsion, while for the HW potential there is the bound state with the
energy 0.01 MeV relative to the $K^{-}p$ +$K^{-}$ threshold. Thus, although
the $\overset{\_}{K}N$ with $I=1$ is attractive, the attraction is not
strong enough to overcome the $K^{-}K^{-}$ repulsion. For the width within
the method of HH we obtain 58.6~MeV and 41.6~MeV with the AY and HW
potentials, respectively. Our results for the binding energy of the $%
K^{-}K^{-}p$ system obtained by the method of HH\ are in reasonable
agreement with calculations obtained using a variational method \cite%
{KanadaJido} and the Faddeev calculations \cite{FadYa4body}. \vspace{-0.1cm}

\begin{table}[tbp]
\caption{ The bound state energies of $K^{-}p$ ($E_{2}$) and $K^{-}K^{-}p$ ($%
B$) systems, and the root-mean-square radius of the $K^{-}$ distribution. $%
\Delta E$ is the binding energy measured from the two-body threshold. }
\label{tab2}\centering
\begin{tabular}{cccccc}
\hline\hline
$K^{-}K^{-}$ & $K^{-}p$ & $<r^{2}>^{1/2},$ fm & $E_{2},$ MeV & $B$, MeV & $%
\Delta E,$ MeV \\ \hline
AY & AY & 1.36 & 30.0 & 31.7 & 1.7 \\ 
$V_{K^{-}K^{-}}=0$ & AY &  &  & 35.3 & 5.3 \\ \hline
HW & HW & 1.96 & 11.42 & 11.43 & 0.01 \\ 
$V_{K^{-}K^{-}}=0$ & HW &  &  & 12.21 & 0.79 \\ \hline\hline
\end{tabular}%
\end{table}

\subsection{\protect\bigskip $KK\protect\overset{\_}{K}$ system}

Recently, there has been increased interest in few-body systems constituted
by two or more kaons. Particularly noteworthy is the possibility of
formation of the quasibound states in a $KK\overset{\_}{K}$ system. We study
the $KK\overset{\_}{K}$ system using a nonrelativistic potential model in
the framework of the method of HH in momentum representation and consider
the $KK\overset{\_}{K}$ system as three interacting kaons. Once the two-body
interactions for the $K\overset{\_}{K}$ and $KK$ subsystems are determined
one can determine the wave function of the $KK\overset{\_}{K}$ system by
solving the Schr\"{o}dinger equation for the Hamiltonian $H=\widehat{T}%
_{3}+V_{\overset{\_}{K}K}(r_{12})+V_{KK}(r_{13})+V_{\overset{\_}{K}%
K}(r_{23}),$ where the potential energy is the sum of the effective $KK$ and 
$K\overset{\_}{K}$ interactions that are the functions of the interparticle
distances $r_{ij}$. For the description of the effective kaon-kaon
interactions we use the local potentials from Refs. \cite{KanadaJido} and 
\cite{JidoKanada} that can be written in one-range Gaussian form (\ref%
{KNpotential}). The set of values of the potential depth $U_{A}^{I}$ for
each \ interaction is given in Refs. \cite{KanadaJido, JidoKanada} and the
range parameter $b$ is chosen to be the same for $KK$ and $KK^{-}$
interactions. We choose two optimized values for the range parameter: 
$b=0.47$ fm (set A) and $b=0.66 $ fm (set B). The strength of strongly
attractive $s- $wave $K\overset{\_}{K}$ interactions was assumed to be the
same for the isospin $I=0$ and isospin $I=1,$ while the strength of the $s-$%
wave $KK$ interaction for the isospin $I=0$ is $U_{KK}^{I=0}=0$ due to Bose
statistics and we consider a weak repulsion for the isospin $I=1$. In Ref. 
\cite{KanadaJido} the $K\overset{\_}{K}$ \ interaction is derived under the
assumption that $K\overset{\_}{K}$ forms the quasibound states $f_{0}$(980)
and $a_{0}$(980) in $I=0$ and $I=1$ channel, respectively, and it reproduces
the masses and widths of these resonances. The strength of the repulsive $KK$
interaction in $I=1$ was fixed to reproduce a lattice QCD calculation \cite%
{Beane} for the scattering length $a_{K^{+}K^{+}}$ $=-0.14$~fm, and a weaker
repulsion that corresponds to the scattering length $a_{K^{+}K^{+}}$ $=-0.10$%
~fm. Results of calculations for the set of potentials A when the $KK$
interaction reproduces the scattering lengths $a_{K^{+}K^{+}}=-0.14$ fm and $%
a_{K^{+}K^{+}}=-0.10$ fm are denoted as A1 and A2, respectively.
Correspondingly, the set of potentials B\textbf{\ }that reproduces those
different scattering lengths hereafter we refer as B1 and B2. The solution
of a system of coupled integral equations for the hyperradial functions
allows us to construct the wave function $\Psi $ for the $KK\overset{\_}{K}$
system and to determine the binding energy $B$. A reasonable convergence for
the ground state energy is reached for the grand angular momentum $\mu
_{\max }=$10 and we limit our considerations to this value. The width of the
bound state is evaluated from the imaginary part of the $K\overset{\_}{K}$
interactions as $\Gamma =-$2 $\left\langle \Psi \left\vert \text{Im}\left(
V_{\overset{\_}{K}K}(r_{12})+V_{\overset{\_}{K}K}(r_{23})\right) \right\vert
\Psi \right\rangle .$ The results of our calculations for the binding energy
and the width for the $KK\overset{\_}{K}$ system along with the results
obtained with a coupled-channel approach based on the solution of the
Faddeev equations in momentum representation and the variational method \cite%
{MartJidoKanada} are presented in Table \ref{tab3}. The total mass the $KK%
\overset{\_}{K}$ system ranges from 1463.4 to 1469.4 MeV when we consider
the same $K$ meson mass $m_{K}=496$ MeV as in Ref. \cite{MartJidoKanada}.
The width falls within the 82- 96 MeV range for all sets of the $K\overset{\_%
}{K}$ and $KK$\ interactions. The quasibound state for the $KK\overset{\_}{K}
$ with spin-parity $0^{-}$ and total isospin 1/2 is found to be below the
three-kaon threshold. The comparison of our results with the results
obtained with the variational method \cite{MartJidoKanada} shows that while
the binding energy found within the HH and variational calculations are
close enough, the difference for the width is more than 20\%. The
alternative scenario is observed for the HH and the Faddeev calculations in
the momentum representation: the difference in masses is more than 40 MeV.
We also perform calculations for the $KK\overset{\_}{K}$ system using $s-$%
wave two-body separable potentials with Yamaguchi form factors from Ref. 
\cite{AMYpotential} that also used in Faddeev and Faddeev-Yakubovsky
calculations \cite{FadYa4body} for the $K^{-}pp$ and $K^{-}K^{-}pp$ kaonic
clusters. The corresponding results are presented in the last column of
Table \ref{tab3} and are very close to the results obtained using the
effective local kaon-kaon interactions for\ the set B. Thus our calculations
within three body nonrelativistic potential model predict a quasibound state
for the $KK\overset{\_}{K}$ system with mass around 1460 MeV that can be
associated with the $K$(1460) resonance. Our results support the conclusion
obtained through the variational calculations and a coupled-channel approach
using the Faddeev equations that\textit{\ }$K$(1460) could be considered as
a dynamically generated resonance. 
\begin{table}[tbp]
\caption{Results of calculations and comparison with the Faddeev and
variational calculations from Ref. \protect\cite{MartJidoKanada}.}
\label{tab3}\centering
\begin{tabular}{cccccccc}
\hline\hline
& $\text{Faddeev \cite{MartJidoKanada}}$ & $\text{Variational with A1 \cite%
{MartJidoKanada}}$ & A1 & A2 & B1 & B2 & $\text{Separable potential}$ \\ 
\hline
Mass, MeV & $1420$ & $1467$ & $1469.4$ & $1468.2$ & $1464.1$ & $1463.8$ & $%
1463.4$ \\ 
$\text{Width, $\Gamma$, MeV}$ & $50$ & $110$ & $82.2$ & $84.0$ & $96.8$ & $98.2$ & $-$
\\ 
$B,\text{ MeV}$ &  & $21$ & $18.6$ & $19.8$ & $23.9$ & $24.2$ & $24.6$ \\ 
$\sqrt{<r^{2}>},\text{ fm}$ &  & $1.6$ & $1.72$ & $1.65$ & $1.61$ & $1.56$ & 
$1.52$ \\ 
$KK\text{ distance, fm}$ &  & $2.8$ & $3.2$ & $2.92$ & $2.72$ & $2.70$ & $%
2.68$ \\ 
$(KK)-\overset{\_}{K},\text{ fm}$ &  & $1.7$ & $1.78$ & $1.68$ & $1.66$ & $%
1.62$ & $1.60$ \\ 
$K\overset{\_}{K}$ distance$,\text{ fm}$ &  & $1.6$ & $1.68$ & $1.65$ & $%
1.64 $ & $1.58$ & $1.55$ \\ 
$(K\overset{\_}{K})-K\text{ distance, fm}$ &  & $2.6$ & $2.9$ & $2.86$ & $%
2.55$ & $2.50$ & $2.47$ \\ \hline\hline
\end{tabular}%
\end{table}
\bigskip
\bigskip
\bigskip
\bigskip
\subsection{\protect\bigskip $\protect\overset{\_}{K}NNN$ clusters}

Recently Faddeev-Yakubovsky calculations \cite{FadYa4body} were made for the
four-particle $K^{-}ppn$ and $K^{-}K^{-}pp$ kaonic clusters, where the
quasibound states were treated as bound states by employing real $s-$wave
two-body separable potential models for the $\overset{\_}{K}\overset{\_}{K}$
and the $\overset{\_}{K}-$nucleon interactions as well as for the $NN$
interaction. Fully four-body nonrelativistic realistic calculations of $%
\overset{\_}{K}NNN$, and $\overset{\_}{K}\overset{\_}{K}NN$ quasibound
states within the method of HH in configuration space, using realistic $NN$
potentials and subthreshold energy dependent chiral $\overset{\_}{K}N$
interactions, were presented in Ref. \cite{BarneaK}. Giving that below we
present the results of our calculations for the $\overset{\_}{K}NNN$, and $%
\overset{\_}{K}\overset{\_}{K}NN$ quasibound states in the framework of
method of HH in momentum representation using AV18 \cite{ArgonneV18} and M 
\cite{Minesota} $NN$ potentials and AY and HW $\overset{\_}{K}N$
interactions as inputs. To find the binding energies with above mentioned
set of potentials, we solve a system of coupled integral equations for the
hyperradial functions $u_{\mu }^{\lambda }(\varkappa ).$ In the calculations
we limit our consideration with the value $\mu _{max}$ $=10$ getting a
reasonable convergence for the binding energy. Using the wave function
obtained for $\mu _{\max }=10$ the width is evaluated through the expression 
$\Gamma =-2\left\langle \Psi \left\vert \mathrm{Im}v_{\overset{\_}{K}%
N}\right\vert \Psi \right\rangle ,$ \ where $v_{\overset{\_}{K}N}$ sums over
all pairwise $\overset{\_}{K}N$ interactions. In Table \ref{tab4} we present
our results for $\overset{\_}{K}NNN$ cluster that we compare with those
obtained via different methods. The results of our calculations for the
energy and the width show dependence on the $NN$ potentials and on the $%
\overset{\_}{K}N$ interactions. However, this dependence is dramatically
different: for the same $\overset{\_}{K}N$ interaction and different $NN$
potentials the ground state energy and the width change by about $3-15\%$,
while for the same $NN$ potential and different $\overset{\_}{K}N$
interaction the energy changes by a factor of more than 3 and the width
changes by more than twice.

\begin{table}[bbp]
\caption{The binding energy $B$ and width $\Gamma $ for the $\protect\overset%
{\_}{K}NNN$ system calculated in the framework of the method of HH in the
momentum representation for different interactions with results from Refs. 
\protect\cite{DotePRC2004}, \protect\cite{FadYa4body} and \protect\cite%
{BarneaK}. The parity $\protect\pi$ includes the eigen parity of antikaon.}
\label{tab4}\centering
\begin{tabular}{ccccccccccc}
\hline\hline
& $J^{\pi }$ & $T$ &  & AV18+AY & M+AY & AV18+HW & M+HW & \cite{BarneaK} & 
\cite{FadYa4body} & \cite{DotePRC2004} \\ \hline
$K^{-}ppn$ & $\frac{1}{2}^{-}$ & 0 & $B$, MeV & 92.1 & 97.9 & 28.6 & 28.9 & 
29.3 & 69 & 110.3 \\ 
&  &  & $\Gamma $, MeV & 83.4 & 84.1 & 30.3 & 30.8 & 32.9 &  & 21.2 \\ \hline
$K^{-}pnn$ & $\frac{1}{2}^{-}$ & 1 & $B$, MeV & 64.6 & 66.7 & 17.2 & 18.7 & 
18.5 &  &  \\ 
&  &  & $\Gamma $, MeV & 74.2 & 80.4 & 27.1 & 31.4 & 31.0 &  &  \\ \hline
$K^{-}ppp$ & $\frac{3}{2}^{+}$ & 1 & $B$, MeV & 101.9 & 107.6 & 25.8 & 28.1
&  &  & 96.7 \\ 
&  &  & $\Gamma $, MeV & 87.9 & 89.8 & 28.1 & 31.2 &  &  & 12.5 \\ 
\hline\hline
\end{tabular}%
\end{table}

For the comparison let's mention that the authors of Ref. \cite{BarneaK}
obtained 18.5 MeV and 31.0 MeV for the binding energy and the width of the $%
K^{-}ppn$ cluster, while calculation within the Faddeev-Yakubovsky equations 
\cite{FadYa4body} with separable potential models for the $\overset{\_}{K}-$%
nucleon and the nucleon-nucleon interactions leads to very deep ground state
energy $-69$ MeV. The comparison of our results for the $K^{-}ppn$ and $%
K^{-}ppp$ clusters obtained for AV18 $NN$ interaction and HW $\overset{\_}{K}%
N$ interaction with calculations \cite{BarneaK} within the variational HH
method for the AV14 $NN$ interaction and shallow chiral $\overset{\_}{K}N$
interaction shows that they are very close. The predictions \cite%
{DotePRC2004} for the binding energy and the width for the kaonic clusters
studied based on a framework of antisymmetrized molecular dynamics and
employing the T potential as a $NN$ interaction and AY potential as a $%
\overset{\_}{K}N$ interaction are presented in the last column of Table \ref%
{tab4}.

Based on the results of our calculations, we can conclude that the pairwise $%
\overset{\_}{K}N$ interaction plays a major role in the formation of the
kaonic bound state and we found that $\overset{\_}{K}N$ effective
interactions \cite{DHWnp2008, DHW}, based on chiral SU(3) dynamics \cite{HW}
leads to a relatively modest binding for the $K^{-}ppn$ and $K^{-}ppp$
clusters that confirms the calculations in \cite{BarneaK}$.$

\subsection{$\protect\overset{\_}{K}\protect\overset{\_}{K}NN$ cluster}

A decade ago in Ref. \cite{AYPL2004} a deeply bound double kaonic \ $%
K^{-}K^{-}pp$ cluster was predicted to be deeply bound with binding energy
of 117 MeV and width 35 MeV. Barnea, Gal and Liverts \cite{BarneaK} perform
a HH calculation in configuration space for the $K^{-}K^{-}pp$ systems based
on the shallow chiral $\overset{\_}{K}N$ interaction model with the
self-consistent energy dependence taken into account and obtained very
shallow bound states with a binding energy that is substantially smaller
than earlier prediction \cite{AYPL2004}. The same tendency can be observed
from Table \ref{tab4} which presents the results of calculations for the
binding energy and width for the double kaonic $K^{-}K^{-}pp$ system.
However our results with the AY $\overset{\_}{K}N$ interaction are close
enough to the prediction of \cite{AYPL2004} and recent Faddeev-Yakubovsky
calculations \cite{FadYa4body}.

\begin{table}[tbp]
\caption{The binding energy $B$ and width $\Gamma $ for the $K^{-}K^{-}pp$
system calculated in the framework of the method of HH in the momentum
representation for different interactions with results from Refs. 
\protect\cite{AYPL2004}, \protect\cite{FadYa4body} and \protect\cite{BarneaK}%
.}
\label{tab5}\centering
\begin{tabular}{ccccccccc}
\hline\hline
&  & AV18+AY & M+AY & AV18+HW & M+HW & \cite{BarneaK} & \cite{FadYa4body} & 
\cite{AYPL2004} \\ \hline
$K^{-}K^{-}pp$ & $B$, MeV & $91.6$ & $92.7$ & $31.5$ & $31.9$ & $32.1$ & $93$
& $117$ \\ 
& $\Gamma $, MeV & $72.4$ & $73.7$ & $78.1$ & $79.2$ & $80.5$ & $-$ & $35$
\\ \hline\hline
\end{tabular}%
\end{table}

\section{Conclusions}

\bigskip Within the framework of a potential model for the kaonic clusters $%
\overset{\_}{K}NN$, $\overset{\_}{K}\overset{\_}{K}N$, $KK\overset{\_}{K}$ $%
, $ $\overset{\_}{K}NNN$, $\ $and $\overset{\_}{K}\overset{\_}{K}NN$ we
perform nonrelativistic three- and four-body calculations using the method
of hyperspherical harmonics in the momentum representation. We examine how
the binding energy and width of the $K^{-}pp$ cluster depends on different
choices of the $\overset{\_}{K}N$ and $NN$\ interactions. Our consideration
includes the realistic Argonne V14 \cite{ArgonneV14}, the semi-realistic MT 
\cite{MT} and T \cite{Tpotential} potentials as inputs for the $NN$
interaction and we employ the phenomenological AY potential and HW potential
constructed based on chiral SU(3) dynamics, as inputs for the $\overset{\_}{K%
}N$ interaction. For all types of considered $\ NN$ interactions, our
calculations predict deeply bound states for the AY $\overset{\_}{K}N$
interaction and a relatively shallowly bound $K^{-}pp$ cluster for the
effective chiral $\overset{\_}{K}N$ interaction. Moreover, the $K^{-}pp$
cluster is the most strongly quasibound three-body system. The results of
our calculations show that the binding energy of the $K^{-}pp$ system
depends entirely on the ansatz for the $\overset{\_}{K}N$ interaction and
substantially changes when we use the AY and HW $\overset{\_}{K}N$
interaction. In regard to the sensitivity of the binding energy to the
details of the $NN$ potentials in Ref. \cite{HW} \ found that as long as the 
$K^{-}pp$ system is only weakly bound, the dependence on different types of $%
NN$ interactions is weak. In fact, our study confirms this conclusion and,
moreover, shows that the dependence on different types of $NN$ interactions
is also weak if the $K^{-}pp$ system is strongly bound.

The strong AY $\overset{\_}{K}N$ interaction is responsible for the
formation of the $K^{-}K^{-}p$ system and this cluster is still bound even
with a much stronger $\overset{\_}{K}\overset{\_}{K}$ repulsion, while the
HW potential leads to the bound state with energy of only 0.01 MeV relative
to the $K^{-}p+K^{-}$ threshold. The mass (binding energy) of the $KK\overset%
{\_}{K}$ system slightly depends on the sets of parameters that determine $K%
\overset{\_}{K}$ and $KK$\ interactions and the width falls into the 82- 96
MeV range for all sets of these parameters. There is reasonable agreement
between these results, the mass obtained using separable AMY \cite%
{AMYpotential} interactions and the variational calculation \cite%
{MartJidoKanada}. Our results for the $KK\overset{\_}{K}$ system support the
conclusion that\textit{\ }$K$(1460) could be considered as a dynamically
generated resonance.

Based on the results of our calculations for four-particle kaonic systems we
also can conclude that the pairwise $\overset{\_}{K}N$ interaction plays a
major role in the formation of the kaonic bound state and the effective
chiral $\overset{\_}{K}N$ interaction gives relatively modest binding for
the $K^{-}ppn,$ $K^{-}ppp$ and $K^{-}K^{-}pp$ clusters.

All our calculations with the effective chiral $\overset{\_}{K}N$
interaction show that the width is always larger than the binding \ energy.
In some cases the width is more than twice as large as the binding energy.
Only for some four-particle kaonic clusters when the input for the $\overset{%
\_}{K}N$ interaction is the AY potential, the binding energy is larger than
the width. As a consequence, perhaps, we are facing a situation where it is
hard to identify the resonances which would make the experimental
observation challenging.

\section*{Acknowledgements}

This work is supported by the MES Republic of Kazakhstan, the research
project IPS 3106/GF4.

\end{document}